# Modal Frustration and Periodicity Breaking in Artificial Spin Ice

*Robert Puttock,\* Alessandra Manzin, Volker Neu, Felipe Garcia-Sanchez, Alexander Fernandez Scarioni, Hans W. Schumacher, and Olga Kazakova*

Here, an artificial spin ice lattice is introduced that exhibits unique Ising and non-Ising behavior under specific field switching protocols because of the inclusion of coupled nanomagnets into the unit cell. In the Ising regime, a magnetic switching mechanism that generates a uni- or bimodal distribution of states dependent on the alignment of the field is demonstrated with respect to the lattice unit cell. In addition, a method for generating a plethora of randomly distributed energy states across the lattice, consisting of Ising and Landau states, is investigated through magnetic force microscopy and micromagnetic modeling. It is demonstrated that the dispersed energy distribution across the lattice is a result of the intrinsic design and can be finely tuned through control of the incident angle of a critical field. The present manuscript explores a complex frustrated environment beyond the 16-vertex Ising model for the development of novel logic-based technologies.

## 1. Introduction

Frustration, the inability of a system to simultaneously achieve a minimum energy within its constitutive parts, produces reconfigurable behaviors in many natural and synthetic environments as it possesses ground state residual entropy.[1–4] Artificially frustrated systems, that is, those manufactured through nanolithography, are used to replicate atomistic spin arrangements on the nanoscale using isolated magnetic nanoelements with preferred moment orientation dictated by their shape, acting as "macro-spins." The freedom to position nanostructures with high precision has enabled the creation of nanoscale models of atomistic effects including 2D Ising models;[5] Boolean satisfiability;[6] and 2D projections of spin ice,[3,7] called artificial spin ice (ASI). ASI are meta-materials composed of geometrically frustrated arrays of ferromagnetic nanoelements, where the non-zero ground state degeneracy is tailored through the spatial arrangement of the lattice components.[4,8] The magnetization in a four-vertex unit cell conforms to a "two-in two-out" spin degeneracy in its ground states; with excited states and reconfiguration possible by application of external stimuli.[9] In four-vertex junctions the energy configurations are summarized by the 16-vertex model, which describes all possible degenerate configurations when all the spins within each element align along their element long axis to form a single-domain configuration (i.e., Ising state).[10–12] ASI may be used as hardware realizations for complex natural systems;[13–15] and in applied research topics including neural networks,[16] computational logic,[17,18] and reconfigurable magnonics.[19–24] Tuning the magnetic properties of ASI lattices through engineered defects or specialized designs are used to induce either deterministic or stochastic effects.[25–28] In particular, lattice designs incorporating coupled nanomagnets have attracted recent attention to create complex mixtures of interactions and study their resulting collective dynamics.[29–32]

A tunable design is presented that incorporates frustrated vertices and coupled nanomagnets into the same structure. By means of magnetic force microscopy (MFM),[33] we demonstrate a confined modal dispersion of heterogeneous energy states, which is not observed in traditional periodic ASI structures. The lattice of macro-spins is perturbed by application of a critical field along the lattice in-plane hard axis resulting in some Ising macro-spins breaking down into complex chiral multi-domain structures called Landau states (LS). Across the lattice, the periodicity of magnetic states transforms into a complex assortment of Ising and LSs that highly depends on the field-application protocol. Utilizing higher degrees of freedom beyond the

R. Puttock, Prof. O. Kazakova
National Physical Laboratory
Teddington TW11 0LW, UK
E-mail: robb.puttock@npl.co.uk

R. Puttock
Department of Physics
Royal Holloway University of London
Egham Hill, Egham TW20 0EX, UK

Prof. A. Manzin, Prof. F. Garcia-Sanchez
Istituto Nazionale di Ricerca Metrologica
Torino 10135, Italy

Prof. V. Neu
Leibniz Institute for Solid State and Materials Research Dresden
Dresden 01069, Germany

Prof. F. Garcia-Sanchez
Departamento de Física Aplicada
University of Salamanca
Pza de la Merced s/n, Salamanca 37008, Spain

A. Fernandez Scarioni, Prof. H. W. Schumacher
Physikalisch-Technische Bundesanstalt
Braunschweig 38116, Germany

16-vertex model is an exciting concept in applications including operational logic[34] and non-Boolean computing.[35] Although LSs are well known to form in nanoelements of reduced aspect ratio and/or large thicknesses,[36,37] their appearance and effect on the frustrated lattice ground state of ASI have not previously been explored. Their formation in the QH lattice yields randomly distributed magnetic configurations, resulting in a loss of long-range periodicity. A deeper understanding of LS formation is pursued, along with their properties within the lattice and their effects on the energy landscape.

## 2. Results and Discussion

### 2.1. Energy Landscapes

The atomic force micrograph in **Figure 1**ai (inset) introduces the novel ASI lattice design. The lattice, composed of permalloy (Py) nanoelements, has a quasi-hexagonal unit cell with four nanoisland (NI) vertices at each junction, thus it is a hybrid of both the Kagome and square ASI designs. The lattice has shared parallel NIs between each unit cell for complimentary ferro-/antiferromagnetic macro-spin coupling. Three junction shapes are present in the lattice, X and Y shapes in a one-to-two stoichiometric ratio, with one Y rotated 180° (herein referred to as reversed Y [rY]). The distribution of junctions results in modal magnetic configurations dependent on the applied field direction, where modality describes multi-frequency populations. Two modal magnetic states are exhibited in Figure 1ai,aii by MFM images at remanence after applied field ($B$) along the $y$- and $x$-axis, respectively. These field directions coincide with the QH-lattice in-plane easy and hard axes, respectively, because of the oblong unit cell. The in-plane shape anisotropy of the lattice was investigated by micromagnetic modeling and is included in the Supporting Information. When magnetized along the easy axis, all junctions conform to the uniform Type II ($T_2$) configuration (Figure 1ai). This state is defined as unimodal (UM) as it possesses a single distribution of energy states.

When the QH lattice is magnetized along its in-plane hard axis the diagonal NIs magnetically align along the field direction (Figure 1aii), and the magnetic state is conserved upon release of the field. The magnetization in the NIs orthogonal to the applied field does not switch. This results in a bimodal (BM) state where the X-shaped vertices are in the $T_2$ magnetic state and the Y/rY-shaped vertices are in the higher energy Type III ($T_3$) state; illustrated as colored blocks in Figure 1aii (inset), cyan and red/blue, respectively. Colored blocks match the respective schematics in Figure 1b. The resulting magnetic landscape is thus confined into dispersed rows of alike energies, producing low-energy pathways in the X-vertices across the globally higher energy lattice.

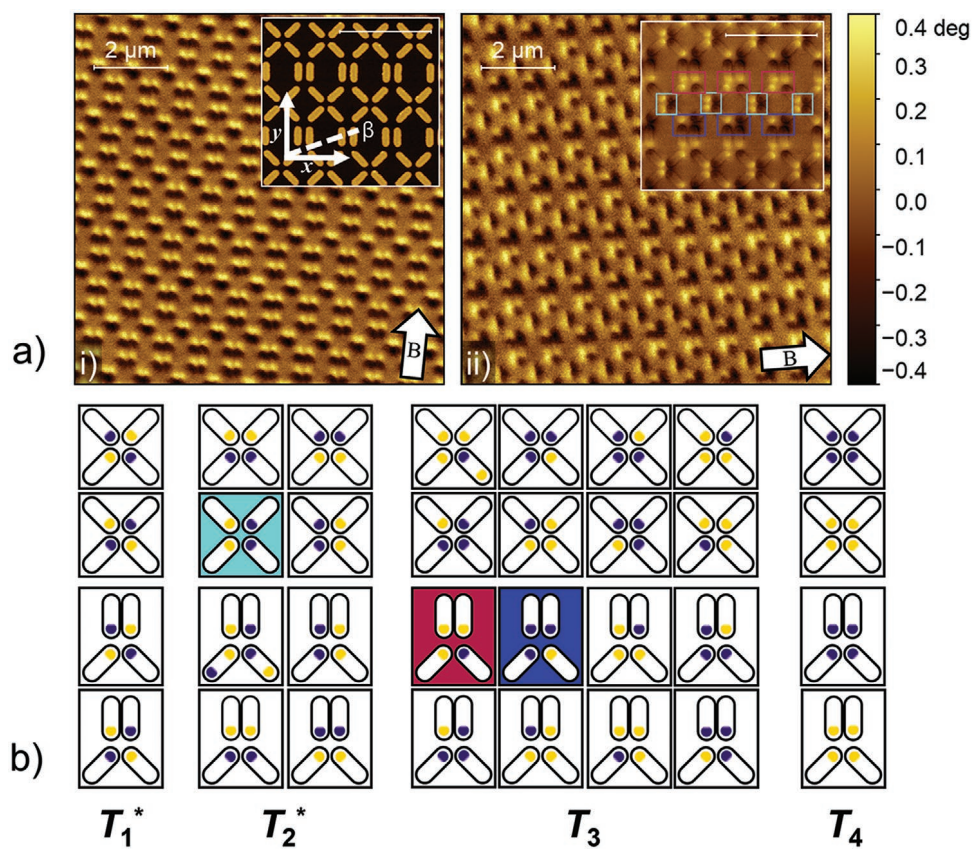

**Figure 1.** a) MFM images at remanence of i) unimodal and ii) bimodal energy configurations for the QH lattice; (i, inset) topography and angular coordinate system for the lattice, and (ii, inset) high resolution MFM image, are included for reference (scale-bar represents 2 μm). b) Schematics of the 16-vertex model for QH-ASI X and rY junctions; four Ising energy types ($T_1$, $T_2$, $T_3$, and $T_4$) are shown where yellow and purple dots represent magnetic charges as seen in MFM images. Colored blocks indicate energy configurations as experimentally observed in a(ii, inset), and those that obey the ice rule are indicated with an asterisk.

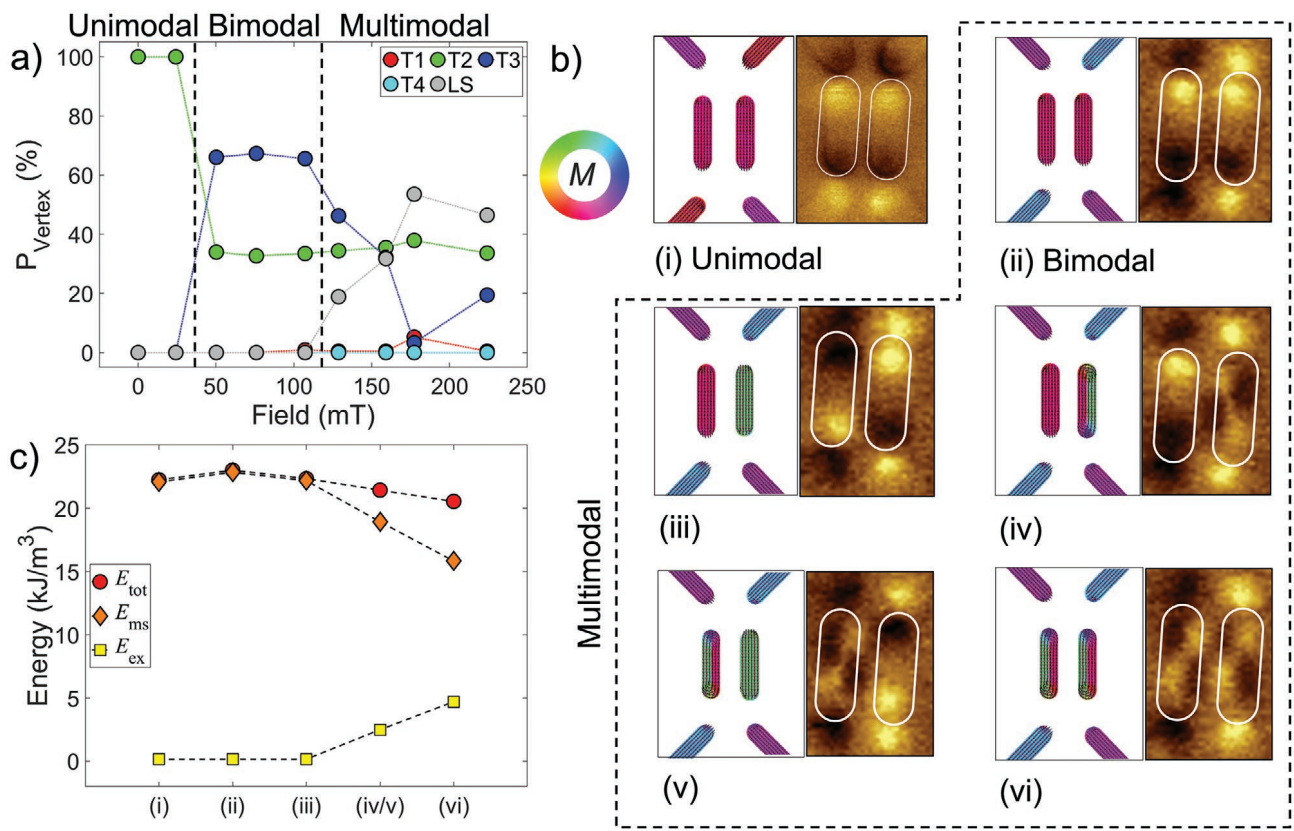

**Figure 2.** a) Vertex population of QH-ASI as a function of field applied along the in-plane hard axis of the lattice after an initial unimodal state (lines are included as guides for the eye). b) Modeled magnetization configurations (the color wheel represents the angle between the magnetization vector in the $xy$-plane and the $x$-axis) and experimental MFM images at remanence (left and right, respectively) of Y/rY vertices in i) unimodal, ii) bimodal, and ii–vi) the five states observed under the multimodal regime. c) Magnetostatic and exchange energy terms (and their summed total) calculated at remanence for the six states displayed in (b).

The evolution of the magnetic states within the lattice was assessed as a function of field magnitude applied along the lattice in-plane hard axis (**Figure 2**a). MFM images of QH ASI were taken at remanence at defined field steps within the range 25–224 mT to assess the transitions between modal landscapes. Initially the lattice magnetization is set in the UM configuration. The MFM images at each field step are displayed in Figure S2a, Supporting Information. Figure 2a plots the population frequency of the energy types across the MFM images as a function of the field, where the plot is divided into three segments to highlight the different modal states. The left and central segments represent the stable field regions where the UM and BM states are formed, respectively. The switch between these modal states at $B = 35$–$50$ mT was not instantaneous, instead it progressed via a two-step mechanism displayed in Figure S2c, Supporting Information.

Considering the same applied-field protocol, the evolution in the magnetic configuration of the QH lattice was studied by micromagnetic modeling, which is reported in **Figure 3**a. The remnant magnetization maps are provided in Figure S3aii, Supporting Information. Here, the remnant BM state appears after applying ≈75 mT field along the $x$-axis direction, closely matching the experimental values. The modeling confirmed that the remanence states at the beginning of the sequence are strongly influenced by the initial UM configuration, thus the lattice exhibits hysteretic properties where the magnetic distribution is frozen into the starting configuration.

Figure 2a was also compared with complementary experimental datasets of the traditional square and alternative-square lattice[38] in Figure S4a,b, Supporting Information. All lattice designs possess a similar switching field between states, but the other geometries remain UM after the transition.

Above a critical field, $B_c \approx 125$ mT, Figure 2a displays a far more chaotic distribution of energy configurations in the third segment. This results from a loss of magnetic periodicity across the lattice as some NI "macro-spins" break down into LSs. This forms a multimodal (MM) state, where more than two vertex energy types occur across the lattice. Figure 2b presents the MFM images and modeled magnetization maps at remanence of the Y/rY junctions are discussed in Figure 2a: the UM (i), BM (ii), and MM states (ii–vi). In the MFM images, the single-domain NIs have uniform magnetization with confined stray-field emanating from the vertices, whereas LSs present as a chequerboard pattern. LSs are not uniformly distributed across the lattice, creating a spatially aperiodic magnetic pattern. From the initial BM Ising state, Figure 2bii, the parallel magnetized NIs can form one of four possible new configurations. Two configurations result from a single NI change: antiparallel distance may still resemble the experimental system well, as scattering events in the lithography process would result in an effective broadening of the nanostructures, reducing the inter-island distance from the original design used as the input to the model.[39] At a higher field (≈300 mT), all of the Ising states in the couple NIs are pushed out of the lattice (Figure 3biv), which is comparable with the 1.0× separation factor in Figure 3aiii. A further discussion of the lattice at high fields is discussed in the following section and in the Supporting Information.

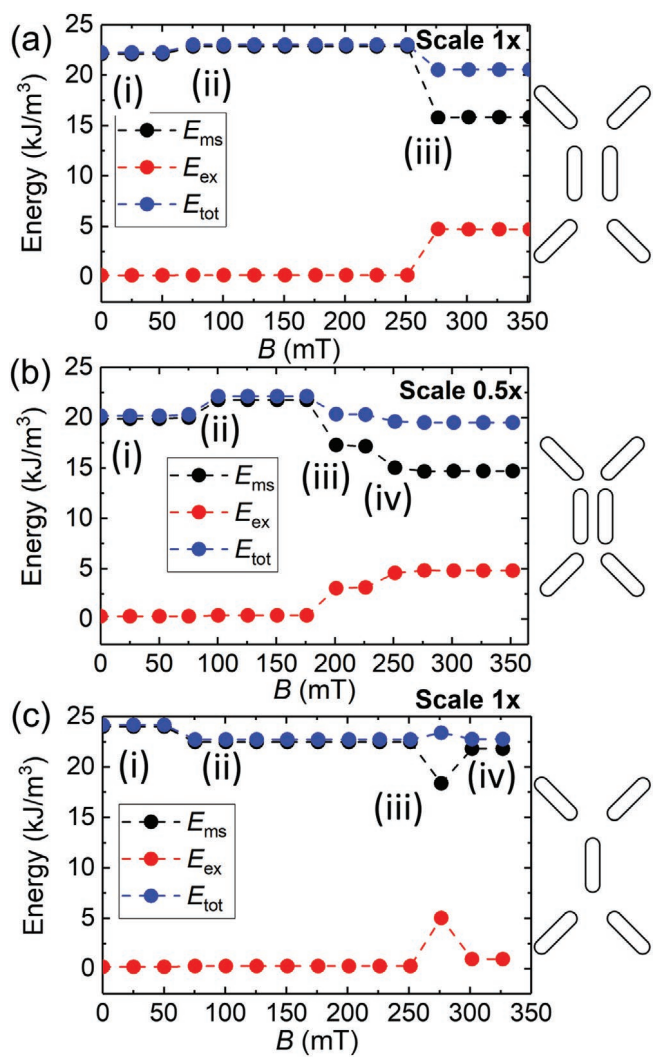

**Figure 3.** Energy terms calculated at remanence for each field step after gradually reaching the field set point and then subsequently releasing the field for the QH-ASI lattice with inter-island separation factor a) 1.0× and b) 0.5× the nominal NI distance c) the predicted energy configuration of the QH-ASI lattice where the parallel islands are replaced by an individual element (with a separation factor of 1.0×). Labeling marks the evolution in the magnetic configuration for the three lattices, which are displayed in Figure S3, Supporting Information.

To understand the complex balance of energies for each configuration in Figure 2b, we extracted the magnetostatic, exchange, and total energies ($E_{ms}$, $E_{ex}$, and $E_{tot}$, respectively) from micromagnetic simulations for the original lattice separation, where each configuration is uniformly distributed across the QH lattice (Figure 2c). For the Py lattice, a saturation magnetization of 860 kA m$^{-1}$, an exchange constant of 13 pJ m$^{-1}$ were used in the simulations. Contributions from thermal noise were assumed to be negligible. For further details of the modeling please see Experimental Section and the Supporting Information. The three Ising configurations (i–iii) are dominated by $E_{ms}$ as the NIs are in a single-domain configuration. $E_{tot}$ for states (i–iii) are 22.25, 23.02, and 22.35 kJ m$^{-3}$, respectively. A sharp increase in $E_{ex}$ is seen upon formation of LSs, however the resulting demagnetization reduces $E_{tot}$ to 21.44 kJ m$^{-3}$ (states iv–v) and 20.55 kJ m$^{-3}$ (state vi). The reduction in $E_{ms}$ upon formation of LSs implies a reduced frustration at the vertex junctions as the LS flux closure occurs within the NI. As a result, the degree of correlation across the lattice is likely diminished, allowing for the loss of long-range order in MFM images and the calculated magnetization maps. Additional modeling of a single NI has been performed and is discussed in Figure S5, Supporting Information. The NI in an LS configuration results in two-fold greater energy than the single-domain case. Therefore, a relative reduction in total energy upon formation of LSs in a lattice demonstrates that it is an energetically stabilized state due to the surrounding magnetic landscape and inter-island coupling.

In order to highlight the unique magnetic properties of the QH lattice, we compared the simulations with a modeled example of the QH design without the coupled NIs (Figure 3c). Magnetization maps and schematics are provided in Figure S3c, Supporting Information. Under the same field history we see that hosting LSs pushes the lattice into a higher total energy (Figure 3ciii), deviating from the cases with parallel NIs as displayed in Figure 3a,b. At the next field increment, LSs are pushed out and the NIs return to an Ising ground state (Figure 3civ). This signals that the presence of the parallel nanomagnets in the QH-ASI lattice causes a deviation from the expected mean-field theory for LS generation within this field protocol, resulting in the MM configuration.

The formation of LSs in ASI lattices provides greater degrees of freedom compared to traditional Ising configurations resulting in a loss of magnetic periodicity. In a lattice without parallel NIs, LS formation is meta-stable resulting from partially saturated spins along the in-plane hard axis of the NI. This effectively creates an unstable "in-between" region where the ice rules are violated. In the presence of coupled NI's in this lattice, the LSs are instead stabilized by the surrounding magnetic interactions, resulting in the favorable summed

Ising pair (iii), and single LS (iv); and two result from both NIs switching: one Ising flip plus one LS (v), and double LS (vi).

In the modeling of the ASI lattice (Figure 3a) the MM configuration is formed at ≈275 mT (Figure 3aiii), where LSs of mixed chirality form uniformly across the lattice (i.e., only state vi). Here, both the field at which the LSs form and the multiplicity of states differ from the experimental dataset. These deviations between the experimental and the modeled values can be mitigated by reducing the inter-island separation (i.e. separation factor) in the model to half of the nominal distance (Figure 3b). Magnetization maps and schematics are provided in Figure S3b, Supporting Information. The greater dipole–dipole coupling between neighboring NIs results in an additional step in the energy progression where the full multitude of states shown in Figure 2b are observed (Figure 3biii). This reduced energy contributions upon their formation. This favorable LS formation and ice-rule violation may apply to other ASI lattices consisting of coupled parallel NIs such as the quadrupolar or trident ASI.[30–32,40]

## 2.2. Landau State Characterization

As we move away from the idealized picture of traditional Ising states in ASI, we further explore the LS formation and characteristics within the QH lattice. In **Figure 4** the distribution in chirality where the flux-closing configuration rotates clockwise (CW) or counterclockwise (CCW) is assessed. Figure 4a depicts MFM images of the square (left) and QH lattice (right) at remanence after applying field $B = 177$ mT at an angle, $\beta = -1.1°$, from both lattice x-axes. The square ASI has a far greater asymmetry in LS chirality (72%) than the QH lattice (53%) and higher proportion of LSs formed (80% and 60% in square and QH lattice, respectively). Figure 4b plots the number of LSs ($N_L$) across MFM images of the QH lattice, normalized by the number of coupled NIs arranged perpendicular to the in-plane field ($N_T$), as a function of $\beta$ across a range of field magnitudes (177–600 mT). The trend follows a Gaussian profile irrespective of the field magnitude, which shows that LS formation is highly dependent on $\beta$. From the fit we can ascertain that the maximum population frequency of LSs in this lattice is 67% in the coupled NIs. A slight x-offset in the peak in Figure 4b likely originates from systematic error in the angle measurement. A full saturation of LS across the lattice from the in-plane field is not possible due to the interactions throughout the surrounding magnetic environment, resulting in a disordered, non-periodic environment of coexisting Ising and Landau states.

Figure 4c plots the angular dependence of the ratio of chiral states for the same dataset as displayed in Figure 4b. Near parity in CW/CCW LSs is demonstrated within ±1° range of the in-plane hard-axis direction. Outside of this range the chirality of LSs is tunable with $\beta$. We also observe a bias toward CCW Landau formation as the field angle deviates from the in-plane hard-axis direction, where the sign of $\beta$ does not affect the chirality bias. We predict this bias is a result of the surrounding magnetic environment upon relaxation of the field. A comparison between the two other structures under study is displayed in Figure S4, Supporting Information.

The experimentally observed proportion and chirality of LSs was compared to the modeled QH lattices in Figure 3 and Figure S3, Supporting Information. At the maximum field magnitude $B = 350$ mT ($\beta = 0$) we see a 100% proportion of calculated LSs in the NIs perpendicular to the incident field for the two QH-ASI with parallel NIs, which does not match the experimental results. However, the chaotic distribution in chirality is matched between the experimental and modeled in these lattices. The state depicted at remanence after a field application of $B = 200$ mT for the reduced inter-island distance in Figure 3biii and Figure S3b(iii), Supporting Information, more closely resembles the final magnetic states observed in the experiments. However, this modeled state is not formed at higher saturation fields such as those depicted in Figure 4b,c. The QH lattice without coupled-nanomagnets (Figure 3c and Figure S3c, Supporting Information) possessed only one chirality of LSs when the saturation field was $B = 275$ mT, which more closely resembles the square lattice's experimental response in Figure 4a and Figure S4, Supporting Information.

This further confirms that the chaotic magnetic configuration is a unique property of the QH lattice when $\beta = 0$ from the presence of the coupled nanomagnets perpendicular to the applied field direction. The ability to tune both LS population and chirality across an ≈6° window results in a degree of control in both formation and magnetic configuration of the lattice, which in turn has far greater use in, for example, logic applications than a truly stochastic effect.

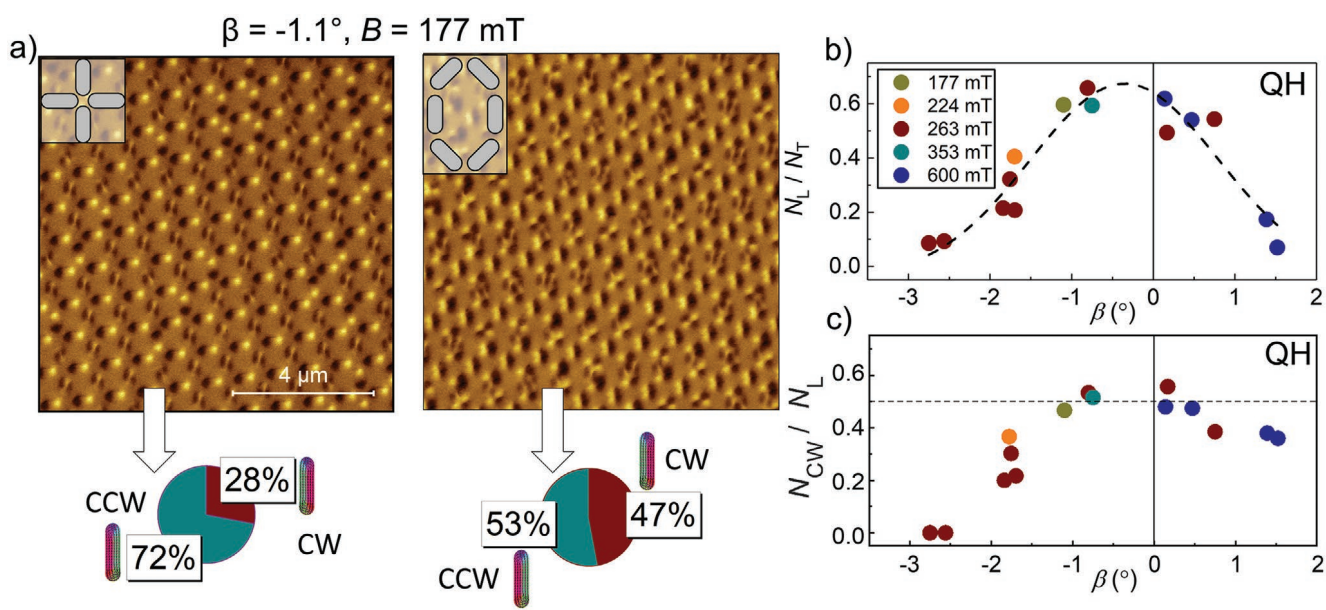

**Figure 4.** a) MFM images of square (left) and QH-ASI (right) lattices at remanence after application of $B = 177$ mT at angle $\beta = -1.1°$ from the x-axis; unit cells are shown in the inset of the respective MFM images. Chirality pie charts depict the distribution of counterclockwise (CCW, teal) and clockwise (CW, burgundy) in the respective MFM images. b) Landau proportion in coupled nanomagnets in the QH lattice as a function of applied field angle ($\beta$) at specified field magnitudes above the critical field. c) Ratio of CW LSs to total number of LSs as a function of $\beta$.

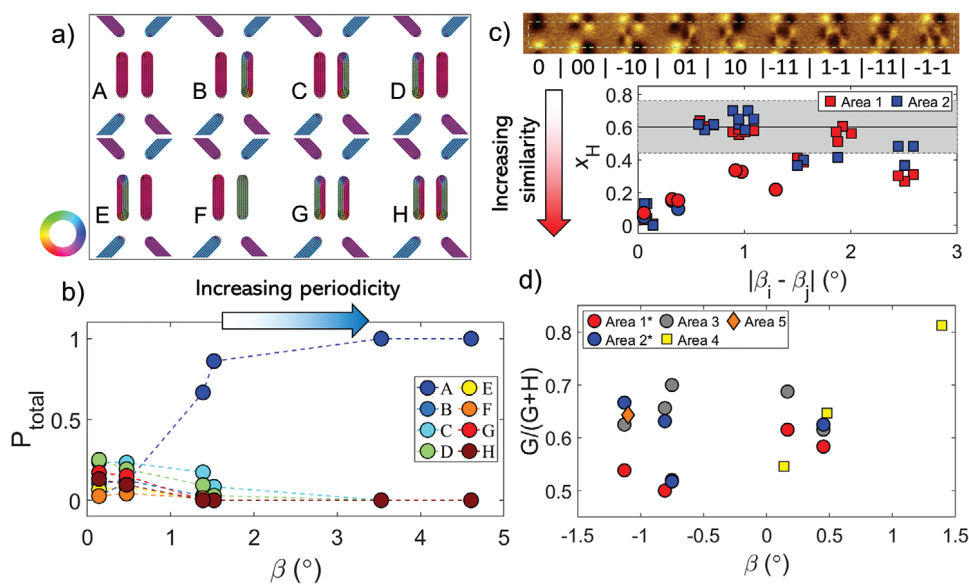

**Figure 5.** a) Modeled magnetization configuration of the expanded Ising and Landau paired states observed in experimental conditions (the color wheel represents the angle between the magnetization vector in the xy-plane and the x-axis). b) Population proportion of states in (a) as a function of applied field angle at $B = 600$ mT. c) (top) Schematic of one line of lattice in multimodal state imaged by MFM, at remanence, after exposure to field $B = 265$ mT along the in-plane hard axis and the assigned numerical integers; (bottom) plot of Hamming distance between MFM images, $i$ and $j$, as a function of absolute difference in the applied field angle, circles are a subset of data where $\beta$ is closely aligned with the lattice in-plane hard axis ($-1.1° < \beta < 0.2°$); grayed area indicates mean and expanded uncertainty from the control dataset. d) Ratio of matching chirality in parallel NIs (state G in (a)) to total number of double-Landau states (G + H in (a)); asterisk indicates areas 1 and 2 from (c); areas-3–5 are randomized-area datasets taken at remanence after application of field with amplitude: 265 mT, 600 mT, and 177 mT, respectively.

To understand further any potential randomness or intrinsic bias in our system, the distribution and chirality of non-Ising states in the NIs under a repeated field protocol is assessed. **Figure 5**a,b depicts the observed states at remanence and their relative population frequencies as a function of $\beta$ ($B = 600$ mT) in MFM images. When $\beta > 3.5°$, only parallel Ising states are observed (state A) in NIs perpendicular to the perturbation field. The other states (B–H) become more prevalent at smaller field angles. Some datasets with energetically equivalent states, "for example", CW/CW and CCW/CCW states, have been combined for simplicity. This plot demonstrates a true break in uniformity across the Y/rY vertices from the formation of multiple iterations of Ising/non-Ising pairs when $\beta \approx 0°$.

To assess and quantify the stochasticity in the MM configurations, identical regions of interest (ROIs) were analyzed by MFM after repeated field application along the lattice in-plane hard axis. Prior to each field iteration on display, the magnetic configuration of the lattice was reset to the UM configuration. MFM images were converted into numerical arrays where NIs with Ising magnetization were assigned to 0, and CW/CCW LSs to 1/–1, respectively (see an example for a row of parallel NIs in Figure 5c, top). The Hamming distance ($x_H$) is a metric to compare differences between two strings of equal size, and it was used to quantify the normalized number of differences in the magnetic configuration between each measurement.[41,42] The $x_H$ value is equal to the sum of differences across the image normalized by the number of elements. Thus, $x_H = 0$ when the magnetic configuration between MFM images $i$ and $j$ is identical, and $x_H = 1$ when the images are completely dissimilar. This quantifies the stochasticity per NI between each MFM image after application of a field with angle $\beta$ for both Ising-LS formation from an Ising configuration, and changes in chirality.

The $x_H$ values obtained in designated ROIs in the array were compared to a dataset where the scan location was taken at random. The non-periodicity of the QH lattice in the MM configuration ensured that the sampled states were dissimilar for each measurement and provided a benchmark value (BMV) for stochastic formation and chirality of LSs. MFM images of the areas #1 and #2 (taken in the corner and central regions of the lattice, respectively) were analyzed separately to determine a possible effect of symmetry breaking due to an absence of neighboring elements. The two ROIs and the control measurement were imaged at remanence after application of a field $B = 265$ mT where the field angle was in the range $-1.8° < \beta < 0.7°$. Figure 4c (bottom) plots $x_H$ against the angular difference of the field vectors between each MFM image. The gray region is the mean BMV and expanded uncertainty ($2\sigma$) from the control dataset, $\bar{x}_H = 0.602 \pm 0.160$. The graph shows that the magnetic configuration in the lattice is generated deterministically as $x_H$ is minimal when the field angles between images are near identical (i.e., $\beta_i - \beta_j \approx 0$). Thus, the individual NIs respond reproducibly to identical field protocols in both the transformation of Ising configuration to LSs, and the LS chirality.

In addition, there is a negligible difference in response when comparing areas #1 and #2, signaling the absence of a symmetry breaking effect. Circular data points in Figure 5c specify MFM images where $\beta$ was closest to the in-plane hard-axis alignment, $-1.1° < \beta < 0.2°$, which is distinctly below the BMV region and provides a linear response as the field angle difference increases. The greatest variability between datasets occurs when $\beta_i - \beta_j \approx 1°$, which is correlated to the steepest regions of the bell-curve in Figure 4b. For a field angle difference greater than $\approx 1.5°$, the Hamming value declines with increasing $\beta_i - \beta_j$, as the lattice is dominated by Ising states.

Highlighting just double LSs, the selectivity of states G and H in Figure 5a is demonstrated in Figure 5d when $\beta = \pm 1.5$. Areas #1 and #2 are the areas described in Figure 5c; whereas area #3 is the control dataset with randomized scan location; areas #4 and #5 are additional datasets in randomized locations obtained at remanence after application of field with amplitude equal to 600 and 177 mT, respectively. The population frequency of G states is independent from $\beta$ within the bounds $-1.2° < \beta < 1°$; instead, the population frequency of G states has far greater variability (between 0.5 and 0.7) compared to the highly reproducible responses shown in Figure 4b,c. This dissimilarity likely stems from the favorable formation of double LSs in the coupled nanomagnets, as it is the lowest energy configuration.

To conclude the results from Figures 4 and 5, the LS state formation and chirality is highly deterministic and dependent on the applied field angle. The complex aperiodic magnetic configuration of the MM state and relative stability of the LSs in the QH lattice is a result of the magnetic environment and the lattice design, specifically in the coupled parallel NIs. In addition, the approximate parity of CW/CCW states is likely a result of balancing the energy of the extended full lattice in a highly correlated system. However, the field angle appears to have little effect on the handedness of the double chiral states in the coupled nanomagnets; contradicting the data from Figures 4b,c. This could be the result of the far reduced $E_{ms}$ upon formation of the double LSs, which in turn reduces the frustration at the vertex junction, and thus the influence of the nearest neighbors for a greater dispersion of chiral states.

## 3. Summary

A violation of the ice rules has been demonstrated in ASI lattices where Ising states break down into LSs in response to a perturbing field protocol. In the novel QH lattice, which exhibits unimodal and bimodal behaviors in the Ising regime, the perturbation field breaks the magnetic periodicity as a plethora of states comprising of Ising and Landau states of mixed chirality are formed. As a result, the typically correlated system loses its long-range order as the frustration at the vertex junctions is impacted. Experimental analysis and micromagnetic modeling demonstrated the LSs form in many ASI lattices, however it is only energetically favorable in a lattice comprised of coupled NIs aligned perpendicular to the perturbation field. This may have significance for other ASI/frustrated designs that are composed of ferro-/antiferromagnetically coupled elements. It has been shown that variations in the perturbation field angle can be used to tune the ratio of both Landau-to-Ising states, and the chirality of LSs. The LS formation and intrinsic chirality is shown to be deterministic and highly repeatable when the field vector is conserved. LS formation not only perturbs the otherwise highly correlated energy landscape of several frustrated ASI lattices, but controllable formation results in additional degrees of freedom for application in frustration-based logic devices.

## 4. Experimental Section

*Fabrication*: ASI nanostructures were patterned using e-beam lithography with a combination of lift-off and Ar ion etching from a sputter-deposited film with a Si/SiO$_x$ (300 nm)/Py (25 nm)/Pt (2 nm) film architecture. The individual NIs were 380 nm × 100 nm laterally in a 100 μm × 100 μm array with a stadium geometry.

*Magnetic Force Microscopy*: MFM measurements were performed using the Bruker Dimension Icon and were conducted at zero field after field application. The MFM probes were NDT-MDT MFM_LM probes. Unless stated otherwise, all MFM images were 10 μm × 10 μm in size. Field angles were calculated from images of the sample between the electromagnet pole pieces and fitted using a purpose-build MATLAB program.

*Micromagnetic Modeling*: Micromagnetic modeling was performed by means of a GPU-parallelized numerical code able to efficiently solve the Landau–Lifshitz–Gilbert equation in large patterned magnetic films.[43,44] The code implemented a geometric time-integration scheme based on Cayley transform for the magnetization update[45] and a fast multipole-based approximation for the evaluation of the magnetostatic field. This was separated into a short-range term, which described the interactions between close NIs via Green integration, and a long-range term, which takes into account the contributions from far NIs via a multipole expansion approximation.[46] The exchange field was calculated with a finite difference method able to handle non-structured meshes and thus suitable for the discretization of NIs with curved boundaries, without introducing fictitious shape anisotropy effects.[47] The damping coefficient was set at 0.1 in order to accelerate the reaching of equilibrium states, according to the analysis reported in ref. [45].

## Supporting Information

Supporting Information is available from the Wiley Online Library or from the author.

## Acknowledgements

This work was supported by project (Grant No. 15SIB06) "NanoMag" in the EMPIR program co-financed by the Euramet participating states and European Union's Horizon 2020 Programme. R.P. and O.K. acknowledge the support of the UK government department for Business, Energy and Industrial Strategy through NMS funding (Low Loss Electronics) and the UK national Quantum Technologies programme. The authors acknowledge S. Gorno for performing some of the MFM measurements in Figure 4b,c and C. Barton for useful discussions.

## Conflict of Interest

The authors declare no conflict of interest.

## Keywords

artificial spin ice, frustrated magnetism, magnetic force microscopy, meta-materials, nanomagnetism

Received: May 19, 2020
Revised: July 31, 2020
Published online: